\documentstyle[12pt,aasms4]{article}

\lefthead{Rubin, Waterman, \& Kenney}
\righthead{Distorted Kinematics among Virgo Galaxies}

\begin{document}

\title{KINEMATIC DISTURBANCES IN OPTICAL ROTATION CURVES AMONG 89 VIRGO DISK
      GALAXIES}

\author{Vera C. Rubin\altaffilmark{1,2}}
\affil{Department of Terrestrial Magnetism, Carnegie Institution of Washington\\
    5241 Broad Branch Road N.W., Washington, D.C. 20015\\
    rubin@gal.ciw.edu}

\author{Andrew H. Waterman\altaffilmark{3}}
\affil{Montgomery Blair High School, Silver Spring, MD}
\and

\author{Jeffrey D. P. Kenney\altaffilmark{1,2}}
\affil{Astronomy Department, P.O. Box 208101, Yale University, New Haven, 
    CT. 06520}

\altaffiltext{1}{Visiting Astronomer, Kitt Peak National Observatory, which 
is operated by AURA, Inc.\ under contract to the National Science
Foundation.} 
\altaffiltext{2}{Observations at Palomar Observatory were made as part of a collaborative agreement between the California Institute of Technology and the Carnegie Institution of Washington.} 
\altaffiltext{3}{present address: Stanford University, Stanford, CA.}

\begin{abstract}

For 89 galaxies, mostly spirals, in the Virgo cluster region, we have obtained  optical long-slit major axis
spectra of the ionized gas. We find: (1) One-half of the Virgo galaxies we observed 
have regular rotation patterns, while the other 50\% exhibit kinematic disturbances 
ranging from mild to major. Velocity complexities are generally consistent with 
those resulting from tidal encounters or accretion events. 
Since kinematic disturbances are expected to fade within $\sim$10$^{9}$ years,
many Virgo galaxies have experienced several
significant kinematic disturbances during their 
lifetimes. (2) There is no strong correlation of rotation curve complexity with Hubble type, with galaxy luminosity, with local galaxy density, or with HI deficiency.  (3) A few Virgo galaxies have ionized gas of limited extent, with velocities  exceptionally low for their luminosities.  In these galaxies the gas must be not rotationally supported. (4) There is a remarkable difference in the distribution of galaxy systemic velocity for galaxies with Regular rotation curves  and galaxies with Disturbed rotation curves. Galaxies with 
regular rotation patterns show a flat distribution with velocities ranging from 
V$_{\rm o}$ = $-$300 km s$^{-1}$ to V$_{\rm o}$ = +2500 km s$^{-1}$; galaxies with disturbed 
kinematics have a Gaussian distribution which peaks at 
V$_{\rm o}$ = +1172$\pm$100 km s$^{-1}$, close to the  cluster mean velocity. This distribution is virtually identical to the  distribution of systemic velocity for elliptical galaxies in Virgo.
However, disturbed spirals are less centrally concentrated than the ellipticals and those near the periphery are more likely to have velocities close to the mean
cluster velocity. Spirals with disturbed kinematics are preferentially on radial orbits, which bring them to the cluster core, where tidal
interactions are strong and/or more common. However, because they spend most of their time near apocenter, we observe them near the periphery of the cluster.
Some may be falling into the core for the first time.
These observations suggest that for a non-virialized cluster like Virgo, galaxies may encounter either local (nearby galaxies) or global (cluster related) interactions. These interactions may alter the morphology of the galaxy, and may also play 
a role in driving the Virgo cluster toward dynamical equilibrium.

\end{abstract}

\keywords{galaxies: clusters; individual (Virgo), galaxies: spirals, galaxies: interactions, galaxies: kinematics \& dynamics, galaxies: distances \& redshifts}

\section{INTRODUCTION}

Clusters of galaxies remain a valuable venue for study.
Morphology and evolution of galaxies differ in clusters from the field
(e.g., Hubble \& Humason 1931, Oemler 1974, Dressler et al. 1997).
As the nearest large cluster, the Virgo cluster offers  the opportunity 
for studying details of environmental effects on galaxy kinematics.
Virgo studies began over two centuries ago, when Messier (1784) noted  the concentration of nebulae 
($``$sans etoile") in Virgo. Yet 150 years  passed before Shapley and Ames 
(1932 and references therein) initiated a large-scale study of the properties of galaxies in the cluster region. 
The most  extensive
study of galaxies in the Virgo cluster is due to Binggeli, Sandage 
and Tammann (1985, 1987, BST; called VCC for 
Virgo Cluster Catalogue), who cataloged 2100 galaxies in the Virgo 
region, and identified 1277 as members. Virgo galaxies have also been studied with 
H$\alpha$ images (Kennicutt \& Kent 1983; Koopmann \& Kenney 1998, 1999), integrated HI profiles (Giovanelli \& Haynes 1983; Helou, Hoffman, \& Salpeter 1984; Warmels 1988; Cayatte et al. 1990), CO  (Kenney \& Young 1989),
radio continuum (Kotanyi 1980), and
x-ray (Fabbiano, Kim, \& Trinchieri 1992).

Surveys of Virgo galaxy properties and comparisons 
with non-cluster galaxies provide evidence that within the core ($\approx$6$^{\rm o}$) of the cluster, 
galaxies are subjected to environmental effects, such as
galaxy-cluster and  
galaxy-galaxy tidal interactions, mergers, gas accretion,
collisions, and gas stripping by the intracluster medium (ICM). Many processes operate in clusters (Oemler 1992; Valluri 1993; Moore et al. 1996;  Kenney \& Koopmann 1999; Barton et al. 1999), although we do not yet know which are dominant, nor the precise evolutionary effects on the disturbed galaxies.

The Virgo cluster is not yet in overall dynamical equilibrium, although some galaxy populations within it are more relaxed than others. 
There are significant sub-structures in the cluster, further evidence of a lack of dynamical equilibrium (BST 1985, 1987, Huchra 1885).
While the ellipticals and lenticulars
have a Gaussian distribution of line-of-sight (l-o-s) systemic velocities
indicating that they may be in the process of establishing a relaxed  population,
the spirals have a complex, multi-peaked distribution (Huchra 1985)
suggesting that they may be clumped into discrete subunits (Sandage, Tammann, \& Binggeli (1985). Many of them  may be presently arriving into the cluster  on elongated orbits (Huchra 1985).
Tully \& Shaya (1984) employed a flow model for the 
local supercluster, offering evidence that many spirals 
in the general region of the Virgo cluster
(including the Milky Way) are on orbits
which will carry them into the Virgo cluster, some within a fraction of a Gyr.

Internal galaxy kinematics are sensitive to tidal interactions, yet
there are few previous measures of gravitational interactions within a single cluster population.
Hence, some years ago we embarked on a program to obtain CCD 
optical long-slit spectra  for disk galaxies in the Virgo 
cluster.
From these high spectral resolution emission line spectra, we determine rotation curves and
study the ionized gas kinematics  for S0$-$Sm galaxies.
Our data are well-suited to study irregularities in velocity fields, which trace disturbances in the galaxies' gravitational 
potentials.

Non-axisymmetric and non-bisymmetric features in the velocity fields
are likely due to tidal events, whether
high or low velocity encounters/collisions, mergers, or accretion events.
Tidal interactions produce a non-bisymmetric disturbance when (m/M)(r/R)$^3$ between the perturbee and the perturber are large; (m/M) is the ratio of the masses, and (r/R) is the ratio of the radius of the perturber to the distances between the two galaxies.
Thus galaxy-cluster tidal effects and near 
galaxy-galaxy tidal encounters between galaxies with nearly similar
masses  are more likely to produce non-bisymmetric disturbances. We think that these types of disturbances are responsible for most kinematic irregularities we discuss in this paper.
Since m=2 structure is common in galaxies, the m=2 mode must
be fairly long-lived, and cannot unambiguously
tell us about recent interactions. In contrast, most other non-axisymmetric kinematic disturbances smooth out in several rotation periods. Hence disturbed kinematics identify galaxies in which gravitational disturbances have taken place within the last billion years.

\section{OBSERVATIONS}

In this paper, we present major axis velocity data for 89 Virgo galaxies 
(plus one galaxy with only a minor axis spectrum), rotation curves for 81, 
and minor axis velocities for 42. Galaxy properties and a journal of observations are compiled in Table 1. For an additional 14 galaxies, most of early 
type located in the central dense Markarian chain, emission is undetected or 
possibly detected only in the nucleus (Table 1 notes).  The observed galaxies (mostly spirals) are fairly well distributed across the cluster and its southern extension (Fig. 1), and include  Hubble types E7/S0 to Spec brighter than about 14.5 (Fig. 2). The sample contains 72 NGC galaxies, 13 UGC (Nilson 1973; UGC), and 5 blue compact objects (BST 1985).

Most galaxies in Table 1 are identified as members of the Virgo cluster. Seventy are classified certain members and 11 are possible members (BST). Nine others have more ambiguous designations. NGC 4165 (V$_{\rm o}$=1882 km s$^{-1}$) and UGC 7181 (276 km s$^{-1}$) were called background by BST. Although their  low rotation velocities imply faint absolute magnitudes, they plot within the scatter of Virgo galaxies on a Tully-Fisher plot.  Seven others (NGC 4064, 4651, 4710, 4713, 4808, 4866, and UGC 7249) are located just outside the BST fields. Their velocities range from 637 to 1114 km s$^{-1}$, except NGC 4866 with V$_{\rm o}$=1994 km s$^{-1}$, and suggest that these are in the outskirts of the cluster.  Contamination from non-cluster galaxies is expected to be minimal. 

There are 42 spirals classified Sa through Scd in the VCC, with B(t) $\le$ 13, 
i $\le$ 40$^{\rm o}$. We have observed 39 (93\%) of these.  Including galaxies classified ``possible" members, with the same Hubble type and inclination limits, we have observed 44 of 56 (79\%).
Corresponding numbers (members and possible members) for B(t)$\le$14 are
60 of 91 observed (66\%) and for B(t)$\le$15 are 64 of 124 observed (52\%). 
Fractions are higher if we limit the sample to ``members".

Of the remaining galaxies in our sample, 24 are classified other than Sa-Scd,  5 are outside the VCC area in VCC, and a few have i $\le$ 
40$^{\rm o}$.
 About two-thirds of our sample galaxies have  no previously published optical velocities extending beyond the nucleus. Twenty-three have ionized gas rotation velocities from  Sperandio et al. (1995), and a few more from Sofue et al. (1998). Resolved HI kinematic data exist for fewer than 25 galaxies observed here (Section 3).

Spectra centered near H$\alpha$ were 
obtained with the Palomar 200-inch and 60-inch telescopes, and  Kitt Peak 
4-m telescope (Tables 1, 2)  for 104 galaxies in the Virgo region. Spectral resolution was $\approx$2 or 3 pixels, and hence near 1\AA~(Table 2).  Position angles of the major axes, determined from available images, are generally within a degree or two of those in the UGC.  For a few galaxies with uncertain position angles, spectra were taken
in several position angles. Slit widths of 1.3$''$ to 2$''$ were chosen to match the instrumental resolution and seeing.  A few large galaxies were observed in two frames. Major axis 
spectra are of high quality, but minor axis spectra taken during intervals of poor weather or bright moon
are of lower quality. For each telescope plus CCD detector combination, 
normal processing procedures were followed. In spectra with a strong galaxy continuum, a mean continuum (of the near H$\alpha$ region)  was removed before measuring.

Details of the velocity measuring procedure are given elsewhere (Rubin, Hunter, \& Ford 1991). Nightsky OH lines are used for the 
two-dimensional wavelength calibration and velocity
zero-point.
At successive  distances from the nucleus along the major axis, velocities of both H$\alpha$ and [NII] (occasionally [SII]) are found from the 
centroid of each emission line. This procedure produces a good estimate of the l-o-s gas velocity at a given position, except near the central few arcseconds of those galaxies with steep velocity gradients.

Several galaxies of kinematic interest have previously been discussed in  detail:  NGC 4550 for its counterrotating stellar disks 
(Rubin, Graham, \& Kenney 1992); the highly disturbed spiral NGC~4438 and its apparent companion
NGC 4435, which may have experienced a high-velocity collision (Kenney et al. 1995); 
the peculiar merger remnant NGC 4424 (Kenney et al. 1996); 
the spiral NGC~4522, which is experiencing ICM-ISM stripping
(Kenney \& Koopmann, 1999); and  
galaxies with kinematically distinct circumnuclear disks (Rubin, Kenney, \& Young 1997). Major and minor axis velocities are 
included here; velocities in other position angles are contained in the references.

Heliocentric systemic velocity as a function of galaxy right ascension is shown in Fig. 3. The mean heliocentric cluster velocity, 1150$\pm$51 km s$^{-1}$ (Huchra  1988) or 1050$\pm$35 km s$^{-1}$ (Binggeli, Popescu, \& Tammann 1993), is poorly defined by our predominantly spiral sample, a recurrent problem with Virgo studies.   
The high systemic velocity dispersion is apparent (Fig. 3) throughout much of the cluster. We adopt M87 as the cluster center, based on ROSAT x-ray observations (Bohringer et al. 1994)

\section{ROTATION VELOCITIES AND CENTRAL VELOCITIES}

Major axis velocities as a function of radius are shown  in Fig. 4.  Mean velocities, for all measures within a small radial bin, are shown along with their 1$\sigma$ errors. For 10 galaxies observed in 1987  (U7181, N4165, U7259, N4237, N4383, N4420, U7590, U7676, N4584, U7932), the position angle of the slit is known, but the algorithm for determining the direction (e.g., NW or SE)
of one end of the slit has been lost. For these, no orientations are included in Fig. 4.	Note that the outermost point for NGC 4519 comes from a companion galaxy, 2.2$'$ NW 
along the major axis (see images, Sandage \& Bedke 1994; Koopmann, Kenney, \& Young 1999). 
Minor axis velocities with respect to the galaxy systemic velocity are shown in Fig. 5.  For NGC 4651, the observed position angle is displaced 9$^{\rm o}$ from the adopted minor axis. For UGC 7623 we have only a minor axis spectrum. We discuss the forms and amplitudes of the rotation curves  in Sections 5 and 6 below.

Representative nuclear spectra (Fig. 6)  illustrate the range in emission line properties, for galaxies ranging from types Amorphous to Sb to dE?, and  blue magnitudes from 10.8 to 18.  NGC 4548 is a Liner which shows the less common
 intensity ratio H$\alpha$/[NII] $\le$ 1; NGC 4639 is a broad lined Sy 1, while NGC 4388 is a  Sy 1.9. NGC 4694 exhibits a typical emission line spectrum, with H$\alpha$ intensity greater than [NII];   and VCC 1941 has  H$\alpha$/[NII] $\ge$ 10. Its strong [SII] lines, relative to virtually absent [NII] lines, identify it as a very low luminosity object (Rubin, Ford, \& Whitmore 1984), consistent with its absolute magnitude M = $-$13 at the cluster distance.
The four NGC objects are classified as members; V1941, with no previously known velocity, was called  a possible member (BST).  Its velocity here, V$_{\rm o}$=1213 km s$^{-1}$, supports its membership. 

Resolved HI observations (as opposed to integrated HI profiles) exist for only a small fraction of Virgo galaxies.  Guhathakurta et al. (1988), Cayatte et al. (1990), and Warmels (1988)  have published HI velocity fields for about 30 Virgo galaxies in common with our sample, but most are at low resolution, and rotation curves are presented only for a subset. An early comparison (Rubin et al. 1989)  of 8 Virgo galaxies (then in common) showed  agreement between  optical and  HI  velocities only for the 3 galaxies observed at 15$''$ resolution. Four Virgo galaxies have recent HI observations:  N4254, (Phookun, Vogel, \& Mundy, 1993), NGC 4321 (Knapen et al. 1993),  N4532 (Hoffman et al, 1999) , and N4654 (Phookun, \& Mundy, 1995). Only NGC 4532 was not on the Guhathakurta et al. list.

It is disappointing that {\it stellar} rotation velocities exist for only 9 of these galaxies: NGC 4374 (Davies \& Birkinshaw 1988), N4435 (Simien \& Prugniel 1997), N4450 and N4569 (Filmore, Boroson, \& Dressler 1986), 4459 (Peterson 1978), N4477 (Jarvis et al. 1988), N4550 (Rubin, Graham, \& Kenney 1992, Rix et al. 1992), N4579 (Palacios et al. 1997), and N4698 (Corsini et al. 1997), and velocity dispersions for 6. Only for NGC 4435, N4550, and N4698 do measured stellar velocities  extend as far (or farther)  then our gas velocities.  For the others, gas velocities extend beyond stellar measures by factors ranging from 2 to over 10. We cannot draw conclusions from this limited material.

Values of  V$_{\rm o}$, the heliocentric systemic velocity for 90 galaxies, are listed in Table 1.  For galaxies with only nuclear emission velocities, the measured (centroid)  nuclear velocity is accepted as  V$_{\rm o}$. For galaxies with emission 
beyond the nucleus, the choice of systemic velocity takes account of the velocity at the strongest continuum, and the outer symmetry of the resulting curve.  The uncertainty
of a single emission line velocity is generally a few km s$^{-1}$, but larger near the nucleus of a galaxy with a steep nuclear velocity gradient, or where extinction  complicates the choice of center.
Uncertainty in  V$_{\rm o}$ is generally of order 10 km s$^{-1}$, increasing to 15 km s$^{-1}$ for galaxies with complex  velocity patterns.

For 74 galaxies with previously published HI systemic  velocities (de Vaucouleurs et al. 1991; RC3), the agreement is excellent, with  mean value 
(V$_{\rm o}-$V$_{21}$) = $-$2.0 km s$^{-1}$, and mean of the absolute values
of (V$_{\rm o}-$V$_{21}$) = 14 km s$^{-1}$.  Five additional galaxies have RC3 velocities, but we follow Roberts et al. (1991) and do not accept as detections  NGC 4293, 4374, 4506, and 4550.  For example, the 21-cm velocity attributed in RC3 to NGC 4374 is an old measure, unsubstantiated by new, more sensitive observations (Huchtmeier \& Richter 1989); it is 181 km s$^{-1}$ lower than our emission line velocities. The fifth galaxy, NGC 4526, is undetected in HI (Kumar \& Thonnard 1983), but detected in CO
 (Sage \& Wrobel 1989), with a broad velocity range. The agreement of our velocities with RC3 {\it optical} velocities is sometimes poor, (e.g., V$_{\rm o}-$V$_{old}$ $\ge$ 300 km s$^{-1}$ in 2 cases),  due to large uncertainties in older values. 

Rotation velocities, (V$_{\rm obs}-$V$_{\rm o}$)/sin $i$, as a function of nuclear distances  are shown (Fig. 7) for 77 galaxies; the galaxies are arranged according to angular extent of measured emission. We assume circular, planar orbits. To transform from angular to linear distances (upper axis, Fig. 7), we adopt  m$-$M=31.0, so the cluster distance is 15.8 Mpc (Jacoby et al. 1992; Freedman et al. 1994) and 1$''$=77 pc, but this choice is irrelevant to our discussion.

Major axis rotation velocities for all galaxies, and velocities along minor axes and other observed position angles, are available in ASCII form from the AJ data bank. The first page of data is shown in Table 3.  For a few  galaxies with uncertain inclinations, we set i=xx in Table 1,  and  only (V$_{\rm obs}-$V$_{\rm o}$) is tabulated in the file.   

As our paper was nearing completion, the spectacular HST long slit (STIS) spectrum of NGC 4374  (M84, E1) was obtained by Bower et al. (1998).  We compare with some amusement our measured ground based velocities with the HST measures (Fig. 8). Although slit widths differ by a factor of 10 (2.0$''$ vs. 0.2$''$), and exposures correspondingly (600s vs. $\approx$5000s), except for the all-important sub-arcsecond velocity peak, the agreement is satisfactory. This comparison supports the maxim that a factor of ten improvement leads to discovery.

\section{EXTENT OF MEASURED EMISSION}

Success in detecting extended emission depends on the instrument, the weather, the integration time, and the galaxy. For our  spectra, the combination of relatively high spectral resolution and long exposures results in a relatively high S/N, especially for the nuclear and HII regions (Fig. 6). Hence, we believe that the measured radial extent depends  principally on the morphological
type and the environmental history of each galaxy.  

The extent of measured emission as a function of Hubble type is illustrated in Fig. 9. The  distance of the farthest measured point along the major axis, r$_{f}$, relative to the isophotal radius R$_{25}$ (Table 1),  ranges from a low of 0.02 for NGC 4374
 (E1) to 1.40 for UGC 7590 (Sbc). As expected,  the median   value increases, from r$_{f}$/R$_{25} \approx$ 0.3 for S0 galaxies, to $\approx$ 0.7  for Sb and Sc. Note, however, that the 
assigned Hubble type of some galaxies in Virgo is misleading, since many galaxies 
classified Sa are actually disk systems with small bulges and low SFRs 
(Koopmann \& Kenney 1998). This may increase the scatter in Fig. 9.
Galaxies with no detected emission, or only possible 
nuclear emission (Table 1 notes) are included in the plot. Rotation curve forms are discussed in Section 5 below.

To illustrate the combined  effects of Hubble type and distance from the cluster core (Fig. 10), we show the extent of detected emission as a function of projected galaxy distance from M87.
For E and S0 galaxies, emission extents vary from 0 to almost 0.4R$_{25}$, with little dependence upon distance from M87. In contrast, Sa spirals show a large variation with core distance. At   1$^{\rm o}- 3^{\rm o}$ from M87, emission extent varies from 0 to 0.7R$_{25}$ (mean=0.25R$_{25}$). 
In the 3$^{\rm o}- 6^{\rm o}$  zone, emission extent increases to 0.2$-$1.0R$_{25}$ (mean=0.55R$_{25}$). For types Sb-Sd, the emission extents are  lower within 2$^{\rm o}$ of M87, and higher, between 3$^{\rm o}$and 6$^{\rm o}$ 
from M87. Thus the well known property of spirals near the Virgo core to  have stripped outer neutral gas disks 
(Haynes, Giovanelli, \& Chincarini 1984; Warmels 1988; Cayatte et al. 1990)
is seen also in the ionized gas in many Virgo spirals (Section 5 below; Koopmann \& Kenney 1998, 1999). 

\section{THE ROTATION CURVES: FORMS}

From these relatively high accuracy rotation curves for Virgo   spirals, there are three properties which we can study. The simplest procedure is to classify the form of each rotation curve as Regular or Disturbed, and to examine its relation to other galaxy parameters.  This is the primary goal of this paper.  A more difficult parameter to evaluate is Vmax; we do this in the following section.  The most difficult parameter to determine is M, the absolute magnitude of each galaxy, via Vmax and Tully-Fisher relation calibrators; we do not attempt this here. 

As expected for galaxies in regions of high density, many Virgo cluster galaxies exhibit complex velocity curves.   Abnormalities consist  generally of: asymmetrical rotation velocities on the two sides of the major axis, falling outer velocities, inner velocity peculiarities, dips in the rotation velocities at intermediate radii, or velocities near zero for the near nuclear gas. Although forms are classified by a comparison with the synthetic rotation curves (SRC) from our earlier studies (Thonnard \& Rubin 1981; Rubin et al. 1985),  
Disturbed forms are easily identified from  knowledge of what  normal rotation curves looks like. We review the formation of SRC in Section 6. 

Examples of velocity curves are shown in Fig. 11.
NGC 4419 has a normal rotation pattern for a high luminosity Sa. In contrast, the starburst galaxy NGC 4383 (Amorphous) has an unusually shallow nuclear velocity rise. The H$\alpha$+[NII] image of NGC~4383 shows filaments of ionized gas,
suggesting significant non-circular gas motions
due to the starburst (Koopmann et al. 1999).
The asymmetrical outer major axis velocities in NGC 4567 (Sc) are probably tidally induced, rising steeply only on the NE where its close companion, NGC 4568, is superposed.  NGC 4498 has a relatively normal velocity pattern on the NW but 
not on the SE; it has an irregular morphology with the outer galaxy off-center and warped (Koopmann et al. 1999),
suggesting a strong gravitational interaction.

For 81 galaxies, we have classified the major axis rotation curve form as Regular (38) or Disturbed  (43: Table 4, Fig. 12).  We attempt a quasi-quantitative measure of the irregularities, and call a rotation curve Disturbed  if we detect a distortion of order 15\% or larger in velocity, extending over 1 kpc or more, in comparison with the SRC.  Minor axis velocities help with  difficult decisions. 
If only the outermost velocity is discrepant, the rotation curve is not classified falling or disturbed, for outermost points are notoriously difficult to measure. Rotation curve forms are not assigned for 5  BCD galaxies, and 3 others with few measured points. With this classification, we make the following conclusions.

1. {\it One-half of the (mostly spiral) Virgo cluster galaxies we observed have rotation patterns called Regular, while about 50\% exhibit more complex internal kinematics, which we label Disturbed}. Abnormalities include asymmetrical rotation velocities on the two sides of the major axis, falling outer velocities, inner velocity peculiarities, dips in the rotation velocities at intermediate radii, or velocities all near zero at small radii. 

For example, the Sc galaxies NGC~4567 and NGC~4568 (sometimes called the Siamese Twins),  an apparent pair with similar l-o-s velocities, both show signs of kinematic disturbance.
Although  NGC~4567 is  significantly disturbed,
the larger  NGC~4568 shows peculiar velocities
only in the outermost data points, and thus is
classified as Regular by our criteria.
NGC~4647 (Sc)  and  NGC~4649 (E) 
form an apparent pair with similar velocities.
While NGC~4647 exhibits an asymmetric H$\alpha$ morphology
suggesting a tidal interaction (Koopmann \& Kenney  1999),
its kinematics appear nearly normal.
Enough time  may not have elapsed
for the kinematic disturbance to be maximally manifest.
The largest kinematic disturbances often appear 
after closest approach (Toomre \& Toomre 1972, Moore et al. 1997); galaxies with mild disturbances may be at early stages of their
interactions.

At the opposite end of the interaction time scale are old encounters.
NGC~4550 and NGC~4698, with counterrotating or misaligned stellar
components, have fairly normal rotation curves. 
Each merger probably occured a sufficiently long time ago that
the velocity fields have become regular.

Complex rotation patterns can arise from galaxy-galaxy interactions. Self-consistent N-body models which explore the first pass of two disk galaxies  (Barton, et al. 1999) produce rotation curves with bumps and dips in the mid regions, and outer portions which rise or fall differently on the two sides of the major axis, much like those we observe. But kinematic disturbances from  galaxy-cluster interactions might also produce similar effects.  Models of disk galaxies falling for the first time into the cluster mean tidal field    show (Valluri 1993) show that m=1 (warp) and m=2 (bar and spiral arms) perturbations result.
Although we do not attempt to distinguish galaxy-galaxy from galaxy-cluster effects, it is noteworthy that eight of our sample galaxies are  predicted to fall into Virgo  (Tully and Shaya 1984) with time scales ranging from 0.1 to 2.2 Gyr (median 0.5 Gyr), and seven of these have rotation forms classified as Disturbed. Of these eight, 2 are called cluster members, 4 possible members, and 2 are outside the VCC boundaries (BST).

Gravitationally disturbed galaxies will return to regular, axisymmetric or bisymmetric kinematics 
in a few rotation periods, $\simeq$1 Gyr.
Consequently, significant kinematic disturbances must occur 
at least once per 1 Gyr for galaxies in Virgo.
Most of these galaxies have likely had several
significant tidal encounters during their lifetime.
This conclusion is broadly consistent with the results of simulations
which show frequent galaxy-galaxy tidal encounters
in clusters (Moore et al. 1996, 1997).

While there are few statistics on the fraction of {\it field} spirals with disturbed rotation curves,  the Virgo set  appears unlike those observed in (nominally) field spirals. The galaxies we studied earlier  (Rubin et al. 1980, 1982, 1985) were chosen to have no near neighbors, and to have relatively normal, unbarred morphology. A reexamination of their rotation curves reveals that most (74\%) of them,  11 of 14 Sa, 14 of 22 Sb, and 16 of 21 Sc,  are {\it very} regular.  The remaining curves have minimal disturbances.  Only two or three show the large scale disturbances we see in some of the Virgo galaxies. Hence we feel secure in attributing the kinematic distortions to the cluster environment.

Other statistics of kinematically  disturbed spirals are fragmentary. Based only on {\it morphology} in the  visual, Zaritsky and Rix, (1997) and infrared,  Odewahn (1996), the fraction of disturbed field spirals has been estimated from 15\% to 50\%.
From studies of non-symmetrical integrated HI profiles, Haynes et al. (1998) estimated the fraction of disturbed galaxies to be high, near 50\%. However, HI profiles arise from a integration of the velocity and the HI distributions; resolved HI observations offer more direct information.  
A study by Swaters et al. (1999) based on kinematically resolved HI observations of lopsided galaxies discusses data for only two such galaxies, but also estimates the disturbed fraction to be at least 50\%. 
The fraction of galaxies which are classified as disturbed in  a given study
depends on the criteria for disturbance and the range of galactocentric radii
considered. Since these vary among the studies cited, we cannot
directly compare our disturbed fraction with others.

2. {\it There is no strong correlation of rotation curve complexity with Hubble type, with galaxy luminosity, or with HI deficiency.} Hubble types are fairly equally distributed among galaxies with Regular and Disturbed rotation forms, (Fig 9, except Sb). Both samples contain mostly Sc's, and  equivalent numbers of S0$-$Sa. Luminosities are also similar among both samples. The median apparent magnitudes, B$_{t}^{\rm o}$=11.90 (Disturbed) and B$_{t}^{\rm o}$=11.93 (Regular), are  similar, and correspond to M$_{B}$=$-$19.1 at the adopted distance.

There is no strong correlation of HI deficiency parameter (Giovanelli \& Haynes 1985) with rotation curve forms (Fig. 13). A comparison of this parameter  for 
Regular and Disturbed rotation galaxies clearly shows  
that Disturbed  galaxies are not preferentially HI-deficient.
This indicates that the mechanism which causes 
HI deficiency, presumably ICM-ISM interactions,
is different from the mechanism which causes kinematic disturbances,
presumably tidal interactions.
In contrast, there is a small (and marginally significant)
excess of Disturbed 
rotation galaxies with normal HI emission (i.e., low HI deficiency).
This small excess might arise if is easier to measure disturbed gas 
kinematics in galaxies with extended gas disks (see 5. below).

3. {\it There is no clear correlation of rotation curve classification with local galaxy density}, as tabulated by Tully (1988). There are both Regular (N4647 and 4808) and Disturbed (N4536) galaxies in the regions of lowest 
(0.21-0.46 gal/Mpc$^{3}$) local density, as well as both Regular  (N4456) and Disturbed (N4477) galaxies in regions of the highest local density (4.06 gal/Mpc$^{3}$). 

NGC 4647 is a good illustration of the difficulties of sorting out local density and rotation curve form.  With a Regular rotation curve which shows little or no sign of an interaction, in a region called low density, it has a close companion on the sky, NGC 4649 (S0, which we have not observed). But the Tully density parameter is not especially sensitive to pairs. Furthermore,  the high relative velocities in many galaxy-galaxy interactions makes it likely that any  external culprit responsible  has long since moved away or merged, so that present local density is only weakly related to earlier density when the interaction occurred. 

An examination of the VCC suggests that low velocity interactions within gravitationally bound pairs or groups is not the major cause of the kinematic disturbances. Of the 11 sample galaxies which are likely to be in bound pairs (e.g., companions within 10$'$ = 46 kpc at the adopted distance, velocity differences of less than 300 km $^{-1}$, and apparent brightness differences of less than 3 magnitudes), 4 are classified as 
Disturbed, 7 as Regular. These disturbed galaxies in pairs likely owe
their disturbances to low-velocity tidal interactions with their companions.
While there are undoubtedly more physically bound pairs and 
groups in Virgo than these 11 (Ferguson 1992), relaxing the identification
criteria reduces the probability of finding true bound systems.
A comparison of the nearest neighbors on the sky and in velocity
shows no significant difference in the probability of belonging to a 
group or similar-mass pair for the Disturbed and Regular galaxies.
This suggests that low velocity 
galaxy-galaxy interactions are not the only cause of kinematic distortions.  High 
velocity tidal interactions (Moore et al. 1997) or galaxy-cluster interactions 
(Valluri 1993) are probably responsible for 
a significant fraction of the disturbances.

4. {\it The distribution of galaxy systemic velocities is different for galaxies with Regular rotation curves and galaxies with Disturbed rotation curves}. 
We show in Fig. 14 the distribution of heliocentric radial velocities  for each class (Table 4).
While galaxies with Regular rotation forms have a relatively flat distribution of observed velocities ranging from $-$300 km s$^{-1}$ to +2600 km s$^{-1}$, galaxies with Disturbed rotation curves have a nearly Gaussian velocity distribution which peaks near the cluster mean velocity. A maximum likelihood fit (Pryor \& Meylan 1993) gives $<$V$>$ = 1172$\pm$100 km s$^{-1}$, and $\sigma_{\rm v}$ = 654$\pm$71 km s$^{-1}$. 

These two different distributions are reminiscent of the distributions which result when galaxies in Virgo are sorted by Hubble type and systemic velocity (Huchra 1985, BST 1987). Early type galaxies show a Gaussian distribution peaked at the cluster velocity ($<$V$>$ = 1200$\pm$46 km s$^{-1}$; $\sigma_{\rm v}$ = 581 (+35/-30) km s$^{-1}$; n=164; Huchra 1985). In contrast, 
late type galaxies have a flatter velocity distribution ($<$V$>$ = 1144$\pm$64 km s$^{-1}$) and consequently higher velocity dispersion ($\sigma_{\rm v}$ = 871 (+49/-42) km s$^{-1}$; n=163; Huchra 1985). In Fig. 15 we superpose the spirals with Disturbed kinematics on the distribution of ellipticals in Virgo (Huchra 1985); the agreement is remarkable. 

While the velocity distribution of the Disturbed spirals 
matches that of the Virgo ellipticals, their spatial distributions are 
quite different, with the ellipticals 
$\sim$30-50\% more centrally concentrated than the Disturbed spirals.
Our new results on the velocity distributions 
offer additional  evidence that galaxies with
Disturbed kinematics have  orbits with large radial (within the cluster) components. In Fig. 16, the  fraction of galaxies with observed velocities farthest from the cluster mean 
(smallest squares) is highest within 3$^{\rm o}$ of M87; here radial orbital motions are aligned close to our l-o-s. With increasing distance from M87, radial orbits make larger angles with the l-o-s, so observed velocities are closer to the cluster mean velocity. The dispersion about the mean cluster 
velocity (adopted as 1100 km/s) decreases from 675 km s$^{-1}$ for Disturbed galaxies observed within 1$^{\rm o}-$2.9$^{\rm o}$ of M87 to 256 km s$^{-1}$ for Disturbed galaxies observed within 5$^{\rm o}-$6.9$^{\rm o}$ of M87. Orbits with large radial components will bring galaxies
close to the cluster center, where tidal interactions with
the cluster and other galaxies are strong and/or more common.
The undisturbed spirals exhibit a significantly less pronounced effect, suggesting that their observed velocity distribution arises from orbits with relatively smaller radial components.

Regardless of the orbital intricacies which produce the observations,  we have been able to identify two populations of Virgo spirals by the characteristics of their rotation patterns. Spirals in the Disturbed population have recently ($\le$1 Gyr) undergone a close encounter with another galaxy or with the cluster tidal field, with consequent kinematic disturbance.  In contrast, the spiral population with normal rotation properties may have suffered no major encounters  within the past $\approx$10$^{9}$ years. One interesting result of this  division of spirals is that five of the six spirals in the sample with negative heliocentric systemic velocities have rotation forms classified Regular.
This is not evidence that they are foreground objects, but rather that they have not had a recent tidal interaction. 

5. {\it Spirals with Disturbed rotation curves are not located preferentially close to M87} (Fig. 16).  As the distance from M87 increases from 1$^{\rm o}$ to 7$^{\rm o}$, in steps of 2$^{\rm o}$, the fraction of galaxies with Disturbed kinematics, relative to Regular,  increases from 0.38  to 0.54 to 0.88. If we include the Sa galaxies without emission in the Disturbed category, the fractions change only to 0.44, 0.54, and 0.88. Although the numbers are small, the relative increase in Disturbed forms, $\Delta$M87$\le7^{\rm o}$, seems fairly secure.

This observation follows as a natural consequence of radial orbits (see 4. above). Galaxies on radial orbits spend most of their 
time near apocenter.
Hence we expect to find Disturbed galaxies 
far from the cluster core, even though the disturbances occur closer to the
cluster core.

We have attempted to examine if there is a bias in our rotation curve classification. The few spirals near the inner core would have truncated gas disks.  Are we more likely to classify an extended rotation curve as Disturbed than we are to classify a truncated curve as Disturbed?   Of the 16 galaxies within 2$^{\rm o}$ of M87, 8 are classified Regular, 7 as Disturbed, and one unclassified. The range of measured gas extents,  r$_{f}$/R$_{25}$, are essentially the same for both types. Of  4 S0s with small inner gas disks,  3 are classified Regular.
Among the galaxies with very extended gas disks,  r$_{f}$/R$_{25} \ge$ 0.75, 6 are classified  Regular, 11 Disturbed. We conclude that any bias in the classification is small, and that the  increasing fraction of Disturbed galaxies at the core periphery is real. 

\section{THE ROTATION CURVES: AMPLITUDES}

Vmax, the maximum amplitude of rotation in the plane of the galaxy, is an important  parameter for  obtaining estimates of absolute magnitudes and distance estimates via the Tully-Fisher  relation (Yasuda, Fukugita, \& Okamura 1997; Federspiel, Tammann, \& Sandage 1998). We list in the final column of Table 1, values of Vmax we have determined. Each value comes from the maximum of the rotation curve, or for 
galaxies with less extended emission, from extrapolating velocities to the R$_{25}$ isophotal radius (Fig. 12).

In evaluating Vmax, we have been guided by the forms of the synthetic rotation curves (SRC). These curves were produced some years ago (Thonnard \& Rubin 1981; Rubin et al. 1985) from our rotation curves of normal (mostly) field spirals,  with systemic velocities ranging from about 1000 km s$^{-1}$ to 8000 km s$^{-1}$, well distributed over the sky.  Rotation curves for galaxies of  Hubble type  Sa (11), or Sb (23), or Sc (20) were used to produce for that Hubble type a sequence of SRC as a function of  absolute magnitude. To determine  absolute magnitudes, each galaxy was placed at the distance inferred from its velocity, for an adopted Hubble constant. Note however, that in making use of the SRC here, we use only their forms; the absolute magnitude scale is not relevant to the discussion.

To review the procedure to form the SRC, we show in Fig. 17 plots  from the 1980 Carnegie yearbook (Thonnard \& Rubin 1981).  The upper panels show the variation of rotational velocity with absolute magnitude for 20 Sc galaxies, each plot for a different value of r/R$_{25}$. Each fitted line is a TF-like relation, but evaluated at a fixed fraction of the galaxy radius, smaller  than  the R$_{25}$ radius. The bottom plot shows the superposition of 5 such fits. To form the SRC for B$_{t}^{o}$ = $-$19 (for example), the velocity at successive values of 0.1 in r/R$_{25}$ is read from the intersections, and a smooth curve drawn. This process is repeated for successive values of B$_{t}^{o}$.

We have no doubt that this is a valid and valuable procedure. But we must point out that the (mostly) field sample of 60 contained 3 Virgo  cluster galaxies,  and these three, NGC 4321 (Sc), 4419 (Sa), and 4698 (Sa), all classified Regular, are in the present sample. As described in those early papers, these three galaxies were placed at the mean Virgo distance. We identify the points for NGC 4321 in Fig. 17. 

A  check on the validity of this procedure  comes from a comparison (Fig. 18) of our Vmax(optical) values with  Vmax(21-cm)  (Federspiel et al. 1998).  These authors correct published values of W$_{20}$ for inclination, turbulent- and z-motions. Our optical values include only inclination corrections and  extrapolations discussed above; our adopted inclinations differ only trivially from those of Federspiel et al. The agreement of the two sets is excellent;
the perpendicular displacement of the points from the line of slope 1 gives $\sigma$(log line width) = 0.037. 

Compared with the 21-cm line width, galaxies with Disturbed rotation curves show no larger scatter than  galaxies with Regular rotation curves. This is evidence that the 21-cm line width and the extrapolated optical rotation velocity measure the same quantity, and the quality of the optical Vmax values are not
compromised by irregularities in the rotation curves. The points  which lie above the line (Fig. 18) have optical velocities (times 2) higher than the 21-cm profile widths. 
The few points which lie highest above the line have observed values of HI deficiency $\ge$0.5. The optical extrapolation procedure may
compensate for the HI truncation, and offer an accurate value of the disk Vmax. 

While it is not the aim of this paper to present a Tully-Fisher analysis nor to determine a distance for the Virgo cluster, we show several TF plots (Fig. 19). Plots (a) and (c) show the correlation of blue apparent magnitude with log Vmax; galaxies are identified by their rotation classification. Galaxies with uncertain values of Vmax are excluded.  Galaxies both Regular and Disturbed are well mixed, and exhibit similar scatter.  

When the galaxies are coded by Hubble type, (Fig. 19b), the characteristic separation of early types toward higher rotation velocity is apparent. When   Hubble types S0$-$Sb are excluded (Fig. 19c),  the scatter is reduced. Although Hubble type dependence disappears for H magnitudes, there are only 18 Virgo galaxies in our sample (Fig. 19d) with H magnitudes (Aaronson et al. 1982). 

A few Virgo galaxies have exceptionally small rotation velocities. Notable are NGC  4064, 4424, 4506, 4584, 4694; all have emission detected only over a small fraction ($\approx0.2R_{25}$) of their radii, with velocities close to zero. Their location on Tully-Fisher plots (Figs. 17, 19) shows that their optical velocities are anomalously low for their optical luminosities, if the l-o-s velocities arise from circular orbits in a circular disk, and if the galaxies are at the distance of the Virgo cluster. While we have no direct measures of their distances, their small gas extents  and peculiar morphologies suggest that they are indeed in the cluster.

They are difficult to classify. Several are called Amorphous or pec, a few are barred, several have no obvious nucleus. N4424 is an apparent merger remnant (Kenney et al. 1996). Perhaps they are en route to becoming spheroidals. A few others, UGC 7171, 7249, and 7784, with low Vmax but with more normally extended gas, define the low-luminosity end of the Tully-Fisher plot (Fig. 19). 

We may have detected galaxies in which the gas is not rotationally supported. Such galaxies have been produced in N-body cluster simulations (Moore, Katz, \& Lake 1995). Following several strong encounters, the galaxy loses angular momentum, and ends up as a prolate figure with gas at the center. Hence rotation along the most elongated axis may be small; rotation  along the apparent minor axis can be larger. Both NGC 4064 and 4424 (Kenney et al. 1996) show approximately zero velocities along their major axes, and small but velocity gradients along their minor axes; we have no minor axis spectra for the others in this set.  Such galaxies deserve more study.

\section{CONCLUSIONS}

The primary purpose of this paper is to present optical kinematic information for a significant number of galaxies in the Virgo cluster, and to identify 
kinematic disturbances in these galaxies.
We present evidence that there are two populations of spirals in the cluster, distinguished by their internal kinematics. Those with Regular rotation properties exhibit a distribution of systemic velocities which is flat, V$_{\rm o}$ = $-$300 km s$^{-1}$ to V$_{\rm o}$ = +2500 km s$^{-1}$.  Those with Disturbed rotation exhibit a Gaussian distribution of systemic velocities which peaks near the cluster systemic velocity, V$_{\rm o}$ = +1172$\pm$100 km s$^{-1}$ and $\sigma_{\rm v}$ = 654$\pm$71 km s$^{-1}$ (Fig. 14,15). Disturbed galaxies are not preferentially near M87, but are prominent near the 6$^{\rm o}$ core periphery. The  Disturbed galaxies probably have orbits with larger radial (relative to the cluster center) components
than the undisturbed galaxies, and some of them
may be falling into the core for the first time. Identifying this population of spirals with disturbed kinematics is the major result of this study. Other conclusions which follow have been discussed in Sections 5 and 6. 

We cannot distinguish between galaxy-galaxy or galaxy-cluster gravitational interactions as the sources which produced these disturbances, although the required perturbations should not be bisymmetric, in order to match the observations. Perhaps there are multiple causes. 
In any cluster,  galaxies will be exposed to both local (galaxy-galaxy) and global (galaxy-cluster) interactions. 
The effects may be more pronounced in a non-virialized 
cluster like Virgo, where tidal effects from substructure
are likely important, and where
infalling galaxies will experience these effects for the first time.

Regardless of the cause, the processes which produce Disturbed rotation curves must also play a role in altering the galaxies' morphologies, and perhaps also in driving the Virgo cluster toward dynamical equilibrium.  In their systemic velocity distribution, the galaxies with Disturbed kinematics resemble the nearly-virialized elliptical population more than they resemble the non-virialized population of spirals with Regular kinematics.  Ellipticals and the Disturbed spirals have similar
distributions of l-o-s orbital velocities, yet the Disturbed spirals are
on average located further from the cluster center. This raises an
interesting question about galaxy morphology and environment.
Do elliptical galaxies owe their morphology to their orbits
within the cluster?

Galaxies we identify as kinematically Disturbed 
are  only that subset presently within the required time
window following the event. 
Interacting galaxies observed at earlier or later  interaction stages 
will have more modest kinematic distortions.
This may explain the lack of simple correlation
between galaxies with kinematic disturbances
and apparent pairs, 
and why other galaxies which may indeed be experiencing
tidal interactions display only minimal
kinematic irregularities.

Thirty years ago, Peebles (1970; see also Lynden-Bell 1967) made a 300-body calculation, matched to velocity and position parameters of galaxies in the Coma cluster. The results confirmed that a gravitationally bound cluster could collapse in an expanding universe, an idea that was then not yet universally accepted. Shectman (1982) extended the calculation to study a more detailed infall model for the formation of the Coma cluster. His conclusion, that a similar calculation ``... should be possible for the Virgo Cluster" is  
now almost 20 years old.  Perhaps its time will soon come. 

With the availability of larger telescopes, observational interests are moving from nearby to very distant clusters. Data from the Virgo cluster should offer a baseline for study of distant clusters, as we attempt to learn when and how galaxies and clusters formed. But the data have relevance to other questions.  They are input for the Tully-Fisher correlation and for our infall to Virgo. They aid in understanding the role that interactions play in the formation and evolution of galaxies within a cluster 
and for the cluster as a whole, and in learning how evolutionary processes differ for field and cluster spirals. Unfortunately, we lack data concerning {\it stellar} orbits within most galaxies. Stellar kinematics can provide valuable evidence for disturbances in gas-poor galaxies, for mergers and accretion events, and also answer questions concerning the similarity of stellar and gas motions. Needed most are three dimensional orbital motions; then we will make real progress in sorting out the kinematic evolution of the cluster.

\acknowledgments

We thank the Directors of Kitt Peak National Observatory and Palomar Observatories for telescope time, and numerous telescope operators for  cheerful assistance. VR thanks Dr. W. Kent Ford, Jr., who participated in early observations of these galaxies, Dr. Paul Schechter for introducing her to the Palomar double spectrograph, Drs. Neta Bahcall, Jay Gallagher, John Graham, Gus Oemler, Richard Larson, and Francois Schweizer for helpful conversations and references, Dr. Allan Sandage for valuable comments on an early draft of this paper, and especially Sandy Keiser for recovering spectra from old 9-track tapes. We thank Drs. J. van Gorkom and D. Burstein for confirming that resolved  21-cm HI observations and H magnitudes are as few as we found. Thoughtful comments from a referee prompted us to enlarge the scope and improve this paper. J.D.P.K. received support from NSF grant AST-9322779, and NASA support to  V. R. and J.D.P.K. comes from HST program GO-5375.  We acknowledge the use of the NASA Extragalactic Database NED operated by IPAC  and
the Observatoire de Lyon Hypercat.

\clearpage

\figcaption{Distribution on the sky for galaxies we have observed in the Virgo cluster region. The X marks the location of M87. ({\it Filled circles}) 89 galaxies for which we have obtained extended velocity measures; ({\it open circles}) 14 galaxies with no
measurable emission. The concentration NW of M87 is the Markarian chain. \label{fig1}}

\figcaption{Hubble type vs total corrected B magnitude (RC3) for Virgo galaxies we have observed.   Filled regions identify galaxies  with no extended emission.
\label{fig2}}

\figcaption{Heliocentric systemic velocity as a function of right ascension of observed galaxies. The X marks the location of M87. ({\it Triangles}) E or S0 galaxies; ({\it filled circles}) Sa and later; ({\it open circles}) galaxies outside of the cluster area surveyed by BST. Note the difficulty in identifying the mean cluster velocity from spiral velocities, the high velocity dispersion over much of the region, and the lower dispersion for the E and S0 galaxies. \label{fig3}}

\figcaption{Major axis l-o-s heliocentric velocities for 89 Virgo galaxies, as a function of nuclear distance. Note the different velocity scales. Each point is the mean of all measures from H$\alpha$ and [NII]$\lambda$6583 within a small radial bin; $\pm$1$\sigma$ velocity errors are shown.
For NGC 4519, the velocity of the companion at r=157$''$ is shown.
For NGC 4550, absorption velocities are plotted as stars. For NGC 4571, velocities in two position angles are plotted separately. Galaxies are sequenced as a function of increasing right ascension, except for VCC compact objects at end. \label{fig4a}}


\figcaption{Minor axis l-o-s velocities minus  systemic velocities,  as a function of nuclear distance. Slit position angles are within 1$^{\rm o}$ of adopted (measured) minor axis, except for NGC 4651, which is displaced 9$^{\rm o}$
from minor axis.
Note the complex velocity variations for many galaxies. \label{fig5}}

\figcaption{High resolution Kitt Peak 4-m nuclear spectra for 5 Virgo
galaxies, arranged by increasing redshift.  For NGC 4639, 4694, and 
4388, counts in the central pixel  (2$''$ x 0.48$''$) are plotted;  
for NGC 4548, the central pixel is 2$''$ x 0.33$''$; and for VCC 1941, the 
sum of the 5 central pixels (2$''$ x 2.4$''$) is shown. VCC 1941 is the faintest Virgo galaxy for which we have velocities.  Emission lines in the Seyfert galaxy NGC 4388 exhibit pronounced asymmetry. Its spectrum 
is plotted displaced 16 Angstroms to the blue, to include the [SII] doublet in the figure. \label{fig6}}

\figcaption{Rotation velocities in the plane of the galaxy, as a function of arcsecs from the nucleus (parsecs on the upper scale), for 77 Virgo galaxies. Galaxies are arranged according to increasing extent of measured velocity. 
For NGC 4519, (last column), the velocity of the companion at 157$''$ is also plotted. Open and filled circles indicate opposite sides of the major axis. See Fig. 12 for velocities in NGC 4424, 4451, 4469, 4506, and 4584. \label{fig7}}

\figcaption{Gas velocities near the nucleus of NGC 4374 (M84, PA=104$^{\rm o}$) 
from HST STIS observations, showing evidence (Bower et al. 1998) for a massive black hole ({\it crosses} and {\it triangles}, where doubled valued). Our 200-inch ground based observations (PA=134$^{\rm o}$, {\it filled circles}) miss the important sub-arcsecond velocity peak, but elsewhere agree  reasonably well. Ground based velocities are heliocentric, with no shift applied. \label{fig8}}

\figcaption{Extent of measured emission, r$_ {f}$, in units of the isophotal radius, R$_{25}$, as a function of Hubble type.  {\it (Open circles)} galaxies with rotation curves classified Regular; {\it (filled circles)} galaxies with rotation curves classified Disturbed.  The median measured emission increases significantly, from r$_{f}$/R$_{25} \approx$ 0.2 for S0 galaxies, to $\approx$ 0.7  for types Sb and Sc. \label{fig9}}

\figcaption{Extent of detected emission, as a function of angular distance from M87, for galaxies of indicated Hubble types. 
The increase (in the mean) in extent of the gas with increasing distance from the cluster core is apparent for  spirals. \label{fig10}}

\figcaption{Rotation velocities for two Virgo Sa galaxies ({\it upper}; NGC 4383 is classified Amorph?, Sandage \& Bedke 1994, and Sa, RC3) and two Sc galaxies ({\it lower}) superposed upon synthetic rotation curves (SRC) formed separately for Sa and Sc (generally field) galaxies (Rubin et al. 1985). Higher velocity curves refer to higher luminosity galaxies. Open and filled circles indicate opposite sides of the major axis.  
We classify the rotation curve form for NGC 4419 as Regular; and Disturbed for NGC 4383 (low velocities near nucleus), NGC 4567 (major axis velocities  asymmetrical at large r), and NGC 4498 (major axis velocities asymmetrical at most radii). \label{fig11}}

\figcaption{Rotation curves for 81 Virgo galaxies, superposed upon synthetic rotation curves (SRC) derived separately for Sa, Sb, and Sc galaxies from our earlier rotation curve studies (see Section 6 for details). Open and filled circles indicate opposite sides of the major axis. Forms are classified Regular (first 38 plots) or Disturbed (last 43) in comparison with SRC. To obtain Vmax, the curves are extrapolated  to r$_{f}$/R$_{25}$ = 1.0, using the appropriate SRC form as a guide. \label{fig12a}}


\figcaption{HI deficiency for spirals with rotation curves classified Regular and Disturbed, showing no strong correlation with rotation curve form, except for a small excess of Disturbed rotation galaxies with low deficiency. See text for details. \label{fig13}}

\figcaption{Histograms showing numbers of Virgo galaxies as a function of heliocentric systemic velocity. ({\it Upper}) galaxies with rotation curve forms judged Regular (Table 4); ({\it lower}) galaxies with rotation forms judged Disturbed. The lower scale gives residual 
galaxy velocity with respect to the nominal Virgo cluster mean velocity, V=1100 km s$^{-1}$.  The dashed regions identify S0 galaxies. Note the nearly Gaussian 
distribution of systemic velocities for galaxies with Disturbed rotation curves, in contrast to the flat distribution of velocities for galaxies with Regular rotation curves.\label{fig14}}

\figcaption{({\it heavy line}) The distribution of 43 spirals (Fig. 14, one S0 included) with Disturbed rotation curves as a function of systemic velocity, superposed on ({\it dashed region}) histogram showing distribution of  164 ellipticals in Virgo (Huchra 1985).
Note the exceptional similarity.\label{fig15}}

\figcaption{Distribution on the sky of galaxies with rotation curves classified Regular ({\it open circles}) and Disturbed ({\it filled squares}). The X marks the location of M87. The largest symbols represent  velocities nearest the cluster mean velocity, i.e., (V$_{\rm o}-<$V$>$) small;  smallest  symbols represents  velocities farthest from from cluster mean. The large circles indicate distances of 1$^{\rm o}$, 3$^{\rm o}$, 5$^{\rm o}$, and 7$^{\rm o}$ from M87. Note that spirals with Disturbed rotation and with velocities near the cluster mean velocity are preferentially observed near the cluster periphery, evidence of their radial orbits. \label{fig16}} 

\figcaption{The variation of rotation velocity with absolute magnitude (Thonnard \& Rubin 1981; reprinted here) for 20 Sc galaxies  studied by Rubin et al. (1980).  Each line in the four upper panels is a TF-like fit to these data, but with velocity read at the indicated fixed fraction of the galaxy radius, smaller  than  the R$_{25}$ radius. The bottom plot shows the superposition of 5 such fits. To form the synthetic rotation curves (SRC) for B$_{t}^{o}$ = $-$19 (for example), the velocity at successive values of 0.1 in r/R$_{25}$ is read from the intersections, and a smooth curve drawn through the points; the process is repeated for successive values of M$_{B}$.   In the present paper, we make use of the forms, and not the magnitudes scales of the SRC. The open circles identify NGC 4321, one of the galaxies in the 1980 study. In the panel for r/R$_{25}$=0.20, we have indicated by x the location for 5 Virgo slow rotators (Section 6), each of which has r$_{\it f}$/R$_{25}$ near 0.20. Note the excess luminosity for their Vmax values.  For these galaxies, the error bars indicate the probable range in M for Virgo distance moduli 30.8$\le$(m-M)$\le$31.8.\label{fig17}}

\figcaption{A comparison of 2 times the maximum rotation velocities for Virgo galaxies found here, with 21-cm width at 20\% of the profiles (Federspiel et al. 1998). The agreement is excellent. The perpendicular displacement of the points from the line of slope 1 gives $\sigma$ = 0.037 magnitudes, which is less than the uncertainties in many values of W$_{20}$ from measured line widths and  inclinations (Federspiel et al. 1998). The median value $\Delta$log(W$_{20}$)=0.047 for the 49 galaxies in common. Values of Vmax are valid even for spirals with kinematic irregularities. \label{fig18}}

\figcaption{Vmax  versus apparent magnitude (TF relation) for Virgo galaxies in this sample; galaxies with uncertain Vmax are excluded. Mean error bars are shown.  ({\it upper}) Galaxies of all Hubble types (a), coded by form of the rotation curve.  Galaxies with low Vmax (e.g., low rotation), are identified with x.  Those with Vmax $\le$ 37 km s$^{-1}$ and r$_{f}$/R$_{25}$ $\approx$0.2 are also shown in Fig 17.  (b) As (a), but Hubble types identified. Note that Regular and Disturbed forms have similar scatter, but that earlier Hubble types have systematically larger Vmax at the same blue magnitude. ({\it lower}) TF plot for galaxies of type Sbc and later (c), showing similar scatter for rotation forms classified 
Regular and Disturbed. (d) TF diagram using H magnitudes available for 18 galaxies in this sample, from Aaronson et al. (1982).\label{fig19}}

\clearpage

\centerline{\bf TABLE CAPTIONS}

\medskip
{\sc TABLE} 1. Parameters and Observing data for Virgo Cluster Galaxies

{\sc TABLE} 2. Observing Parameters

{\sc TABLE} 3. Major Axis Rotation Velocities as a function of Nuclear Distance 

{\sc TABLE} 4. Rotation Curve Forms


\clearpage

\begin{references}

\reference{} Aaronson, M. et al. 1982, ApJS, 50, 241

\reference{} Barton, E. J., Bromley, B. C., \& Geller, M. J., 1999, in press

\reference{} Binggeli, B., Sandage, A., \& Tammann, G.A. 1985, AJ, 90, 1681

\reference{} Binggeli, B., Sandage, A., \& Tammann, G.A. 1987, AJ, 94, 251

\reference{} Binggeli, B., Popescu, C. C., \& Tammann, G. A. 1993, A\&AS, 98, 275

\reference{} Bohringer, H., Briel, U. G., Schwarz, R. A., Voges, W., Hartner, 
     G., \& Trumper, J. 1994, Nature, 368, 828

\reference{} Bower, G. A. et al. 1998, ApJ, 492, L111

\reference{} Cayatte, V., van Gorkom, J. H., Balkowski, C. \& Kotanyi, C.
    1990, AJ, 100, 604

\reference{} Corsini, E. M., Pizzella, A., Bertola, F., \& Beltran, J. C. 
     V. 1997, in {\it Dark and Visible Matter in Galaxies} ed. M. Persic \& 
     P. Salucci, ASP Conference Series \#117  (Salt Lake City: ASP)

\reference{} Davies, R. L., \& Birkinshaw, M. 1988, ApJS, 68, 409

\reference{} de Vaucouleurs, G., de Vaucouleurs, A., Corwin, H. G., Buta, R.
    J., Paturel, G., Fouqu\'{e}, P. 1991, {\it Third Reference Catalog of
    Bright Galaxies}, (New York: Springer-Verlag) (RC3)

\reference{} Dressler, A., Couch, W. J., Smail, I., Ellis, R.
    S., Barger, A., Butcher, H., Poggianti, B. M., \& Sharples, R. M. 1997, 
    ApJ, 490, 577

\reference{} Fabbiano, G., Kim, D.-W., \& Trinchieri, G. 1992, ApJS, 80, 531

\reference{} Federspiel, M., Tammann, G. A., \& Sandage, A. 1998, ApJ 495, 115

\reference{} Fillmore, J. A., Boroson, T. A., \& Dressler, A. 1986, ApJ, 
    302, 208

\reference{} Ferguson, H. C. 1992, MNRAS, 255, 389 1992

\reference{} Freedman, W. L. et al. 1994, Nature, 371, 757

\reference{} Giovanelli, R., \& Haynes, M. P., 1985, ApJ, 292, 404

\reference{} Giovanelli, R., \& Haynes, M. P., 1983, AJ, 88, 881

\reference{} Guhathakurta, P., van Gorkom, J. H., Kotanyi, C. G., \& Balkowski,
     C. 1988, AJ, 96, 851     

\reference{} Haynes,  M. P., Giovanelli, R., \& Chincarini, G. L. 1984, ARAA,
     ed. G. Burbidge, D. Lazer, J. G. Phillips, Vol. 22, p. 445

\reference{} Haynes, M. P., Hogg, D. E.,  Maddalena, R. J., Roberts, M. S., 
     van Zee, L. 1998, AJ, 115, 62

\reference{} Helou, G., Hoffman, G. L., \& Salpeter, E. E. 1984, ApJS, 55, 433

\reference{} Hoffman, G. L., Lu, N. Y., Salpeter, E. E., \& Connell, B. M. 
      1999,  AJ, 117, 811

\reference{} Hubble, E., \& Humason, M. L. 1931, ApJ, 74,  43

\reference{} Huchra, J. P. 1985, in {\it The Virgo Cluster}, ed. O.-G. Richter
     \& B. Binggeli (Garching bei Munchen: ESO), p. 181

\reference{} Huchra, J. P. 1988, in {\it The Extragalactic Distance Scale}, 
    ed. S. van den Bergh \& C. J. Pritchet, ASP Conference Series, Vol. 4 
    (Salt Lake City: ASP), p. 257

\reference{} Huchtmeier, W. K., \& Richter, 0.-G. 1989, A General Catalogue 
     of HI Observations of Galaxies (New York: Springer-Verlag) 

\reference{} Jacoby, G. H., Branch, D., Ciardullo, R., Davies, R. L., Harris,
     W. E., Pierce, M. J., Pritchet, C. J., Tonry, J. L., \& Welch, D. L. 
     1992,  PASP, 104, 599

\reference{} Jarvis, B. J., Dubath, P., Martinet, L., \& Bacon, R. 1988,
     A\&AS, 74, 513

\reference{} Kenney, J. D. P., \& Young, J. S. 1989, ApJ, 344, 171

\reference{} Kenney, J. D. P., \& Koopmann, R. A. 1999, AJ 117, 181

\reference{} Kenney, J. D. P., Rubin, V. C., Planesas, P., \& Young, J. S.
     1995, ApJ, 438, 135

\reference{} Kenney, J. D. P., Koopmann, R. A., Rubin, V. C., \& Young, J. 
     S. 1996, AJ, 111, 152

\reference{} Kennicutt, R. C., \& Kent, S. M. 1983, AJ, 88, 483

\reference{} Knapen, J. H., Cepa, J., Beckman, J. E., Soledad del Rio, M., \&
     Pedlar, A. 1993, 416, 563

\reference{} Koopmann, R. A., \& Kenney, J. D. P. 1998, ApJ, 497, L75

\reference{} Koopmann, R. A., \& Kenney, J. D. P. 1999, in prep.
             
\reference{} Koopmann, R. A., Kenney, J. D. P., \& Young, J. 1999, in prep.

\reference{} Kotanyi, C. G. 1980, A\&AS, 41, 421

\reference{} Kumar, C. K., \& Thonnard, N. 1983, AJ, 88, 260

\reference{} Messier,  Connaissance des Temps, Paris, 1784, 263
     
\reference{} Lynden-Bell, D. 1967, MNRAS, 136, 101

\reference{} Moore, B., Katz, N., \& Lake, G.,  1995, in {\it New Light on 
     Galaxy Evolution} IAU, Vol. 171, ed. R. Bender \& R. L. Davies 
     (Kluwer: Dordrecht), p. 203

\reference{} Moore, B., Katz, N., \& Lake, G.,  1997, ApJ, 495, 139

\reference{} Moore, B., Katz, N., Lake, G., Dressler, A., 
     \& Oemler, A., Jr. 1996, Nature, 379, 613

\reference{} Nilson, P. 1973, Uppsala General Catalogue of Galaxies,
     (Uppsala:Uppsala Offset Center AB)

\reference{} Odewahn, S. 1996, in {\it Barred Galaxies}, ed. R. Buta, D. A. Crocker, \& B. R. Elmegreen, ASP Conference Series Vol. 91, (Salt Lake City: 
     ASP) p. 30

\reference{} Oemler, A., Jr. 1974, ApJ, 194, 1

\reference{} Oemler, A., Jr. 1992, in {\it Clusters and Superclusters 
     of Galaxies}, ed. A. C. Fabian (Dordrecht: Kluwer), p. 29

\reference{} Palacios, J., Garcia-Vargas, M. L., Diaz, A., Terlevich, R., \& Terlevich, E. 1997, AA, 323, 749  

\reference{} Peebles, P. J. E. 1970, AJ, 75, 13

\reference{} Peterson, C. J. 1978, ApJ, 222, 84

\reference{} Phookun, B., Vogel, S. N., \& Mundy, L. G. 1993, ApJ, 418, 113

\reference{} Phookun, B., \& Mundy, L. G. 1995, ApJ, 453, 154


\reference{} Pryor, C., \&  Meylan, G. 1993, in {\it Structure and Dynamics 
     of Globular Clusters}, ed.  G. Meylan \& S. Djorgovski, ASP 
     Conference Series Vol. 50,  (Salt Lake City: ASP) p. 357

\reference{} Rix, H.-W., Franx, M., Fisher, D., \& Illingworth, G. 1992, 
     ApJ, 400, L5

\reference{} Roberts, M. S., Hogg, D. E., Bregman, J. N., Forman, W. R., 
     \& Jones, C. 1991, ApJS, 75, 751

\reference{} Rubin, V. C., Ford, W. K., Jr., \& Whitmore. B. C. 1984, ApJ, 
     281, L24

\reference{} Rubin, V. C., Ford, W. K., Jr., \& Thonnard, N. 1980, ApJ, 238, 471 
\reference{} Rubin, V. C., Ford, W. K., Jr., Thonnard, N. \& Burstein, D. 1982, 
    ApJ 261, 439
\reference{} Rubin, V. C., Graham, J., \& Kenney, J. D. 1992, ApJ, 394, L9

\reference{} Rubin, V. C., Hunter, D. A., \& Ford, W. K., Jr. 1991, ApJS, 
     76, 153

\reference{} Rubin, V. C., Kenney, J. D., Boss, A. P., \& Ford, W. K. Jr., 
     1989, AJ, 98, 1246

\reference{} Rubin, V. C., Kenney, J. D. P., \& Young, J. S. 1997, AJ, 113,
     1250  

\reference{} Rubin, V. C., Burstein, D., Ford, W. K., Jr., \& Thonnard, 
     N. 1985, ApJ, 289, 81

\reference{} Sage, L. S., \& Wrobel, I. M. 1989, ApJ, 344, 204

\reference{} Sandage, A., \& Bedke, J. 1994, {\it The Carnegie Atlas of
   Galaxies},  (Washington: Carnegie Institution of Washington)

\reference{} Sandage, A., Binggeli, B., \& Tammann, G. A. 1985, in {\it The
     Virgo Cluster}, ed. O.-G. Richter
     \& B. Binggeli (Garching bei Munchen: ESO), p. 239

\reference{} Shapley, H., \& Ames, A. 1932, Harvard Ann. 80, 41

\reference{} Shectman, S. A. 1982, ApJ, 262, 9

\reference{} Simien, F., \& Prugniel, Ph. 1997, A\&AS, 126, 15

\reference{} Sofue, Y., Tomika, A., Tutui, Y., Honma, M., \& Takeda, Y. 1998, 
    PASJ, 50, 427

\reference{} Sperandio, M., Chincarini, G., Rampallo, R., \& de Souza, R. 1995,
    A\&AS, 110, 279

\reference{} Swaters, R. A., Schoenmakers, R. H. M., Sancisi, R., \& van Albada, T. S. 1999, MNRAS, 304, 330

\reference{} Thonnard, N. \& Rubin, V. C. 1981, Carnegie Yrb, 80, 551

\reference{} Toomre, A. \& Toomre, J. 1972, ApJ, 178, 623    

\reference{} Tully, R. B. 1988, Nearby Galaxies Catalog, (Cambridge: Cambridge 
    University Press)

\reference{} Tully, R. B., \& Shaya, E. J. 1984, ApJ, 281, 31

\reference{} Valluri, M. 1993, ApJ, 408, 57

\reference{} Warmels, R., 1988, A\&AS, 72, 19

\reference{} Yasuda, N., Fukugita, M., \& Okamura, S. 1997, ApJS, 108, 417

\reference{} Zaritsky, D. \& Rix, H.-W. 1997, ApJ, 477, 118




\end{references}
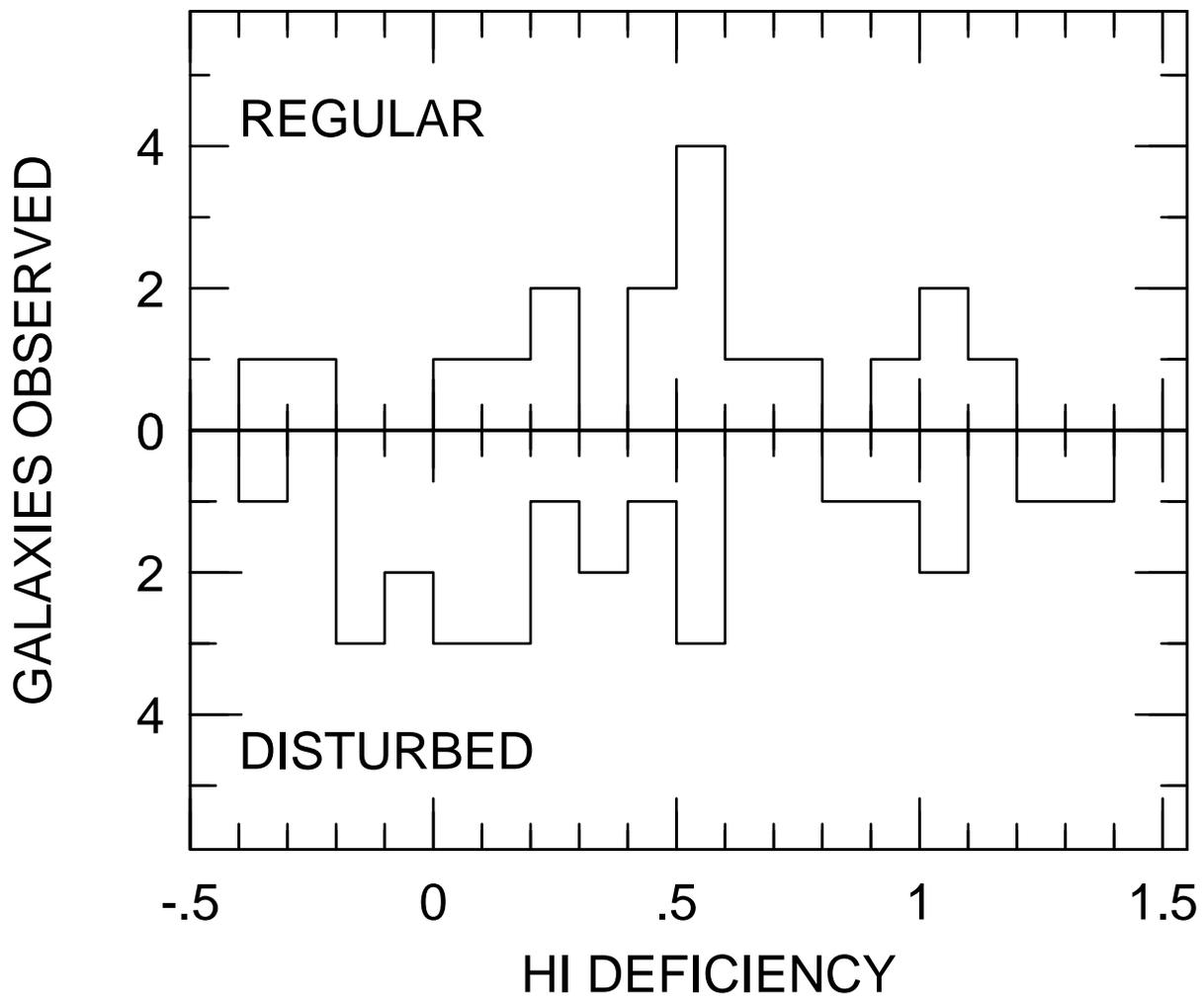
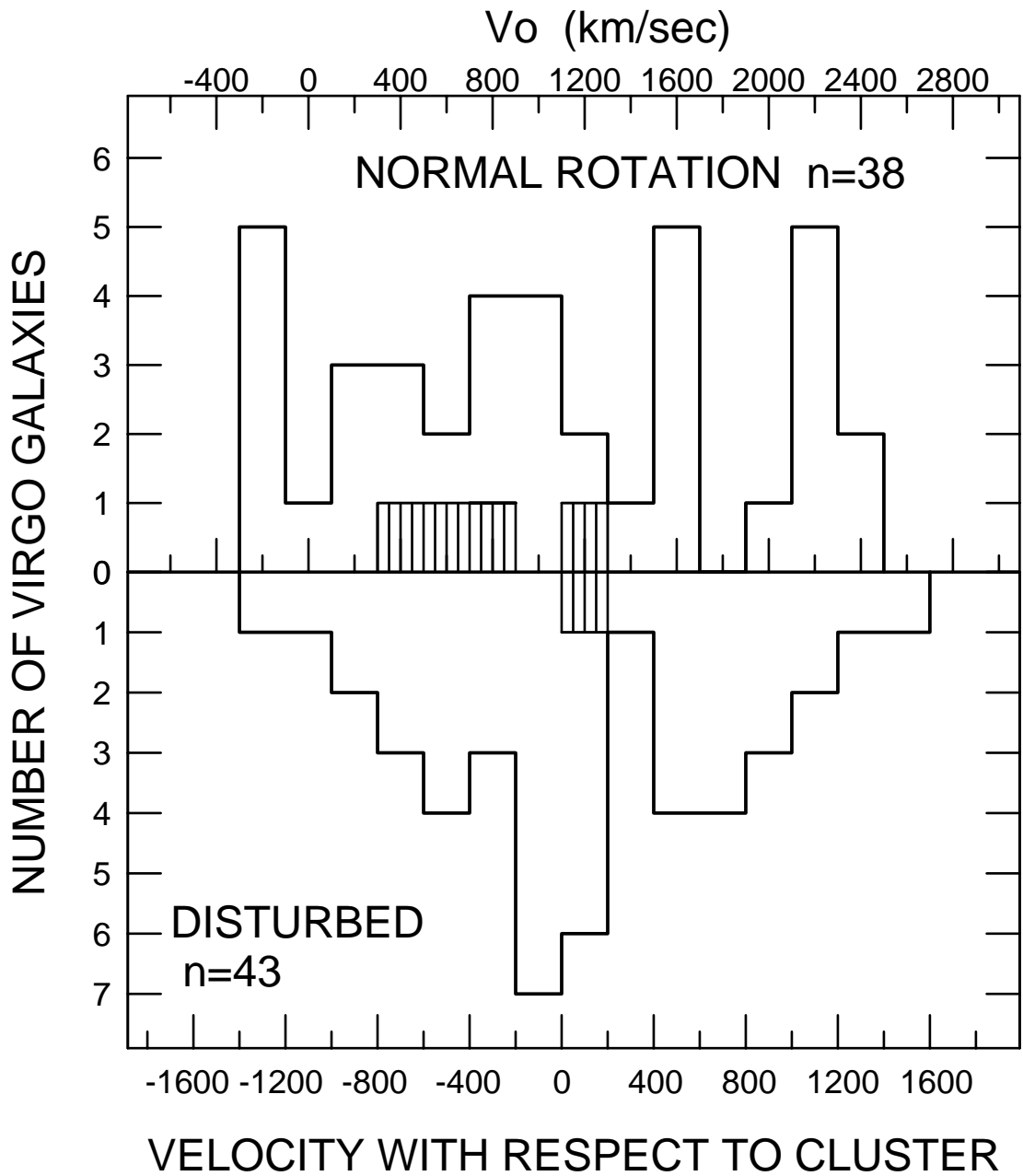
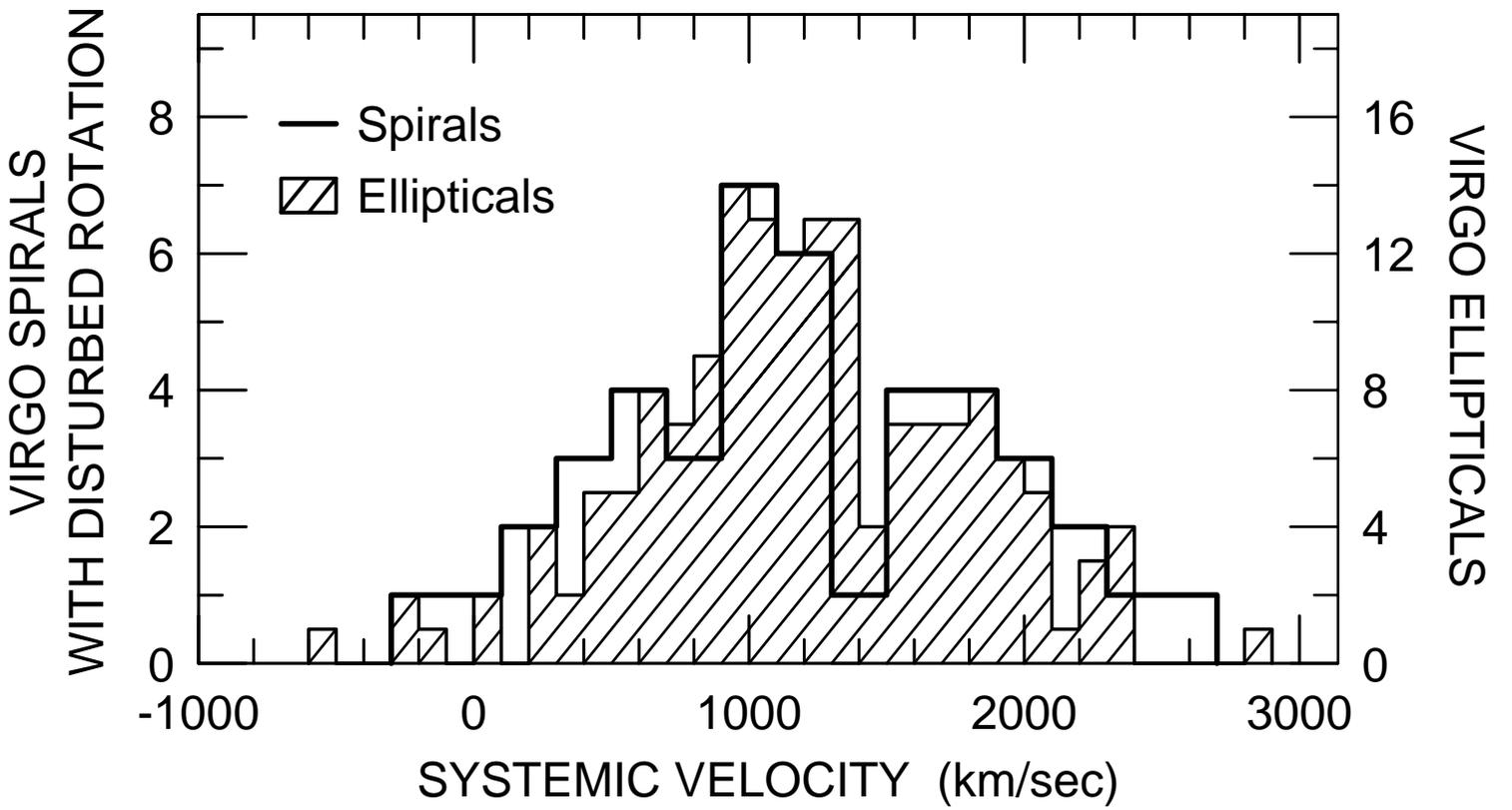
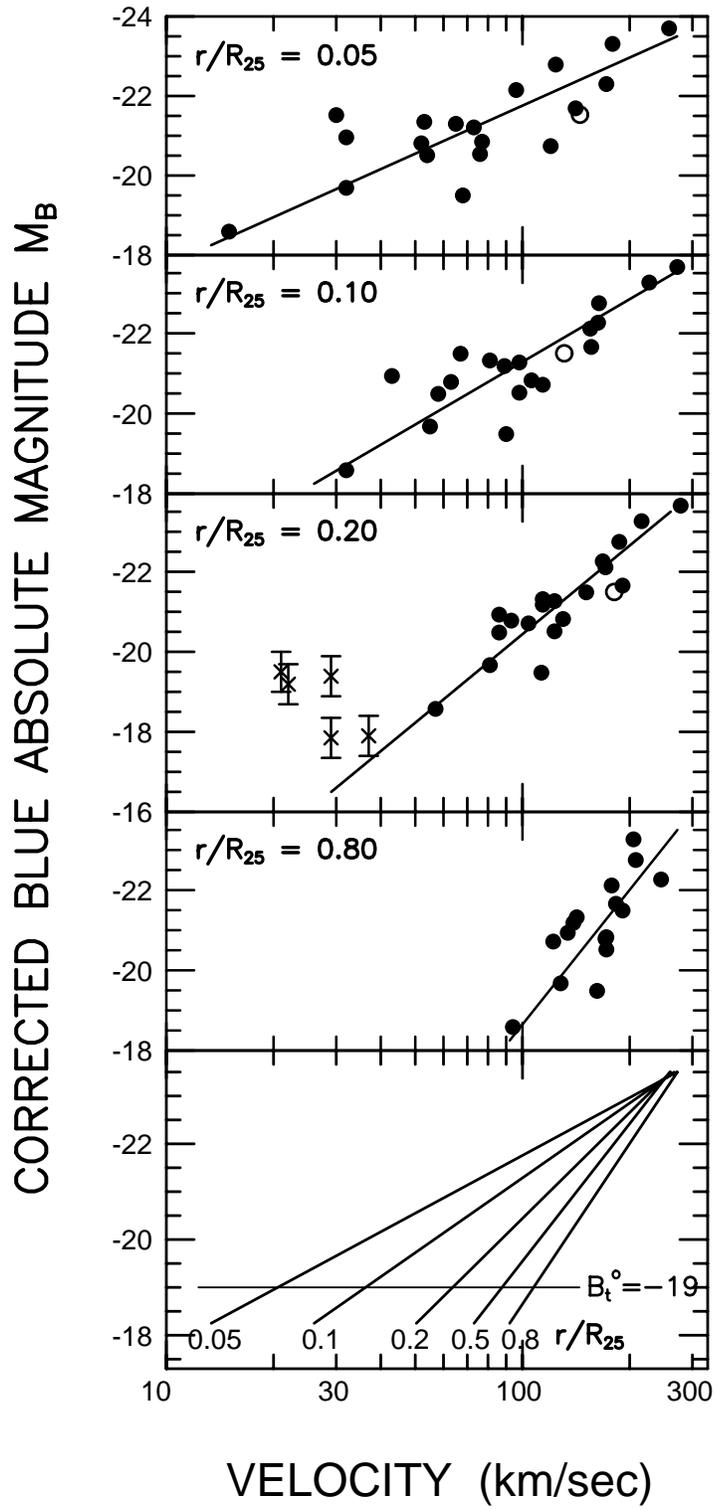
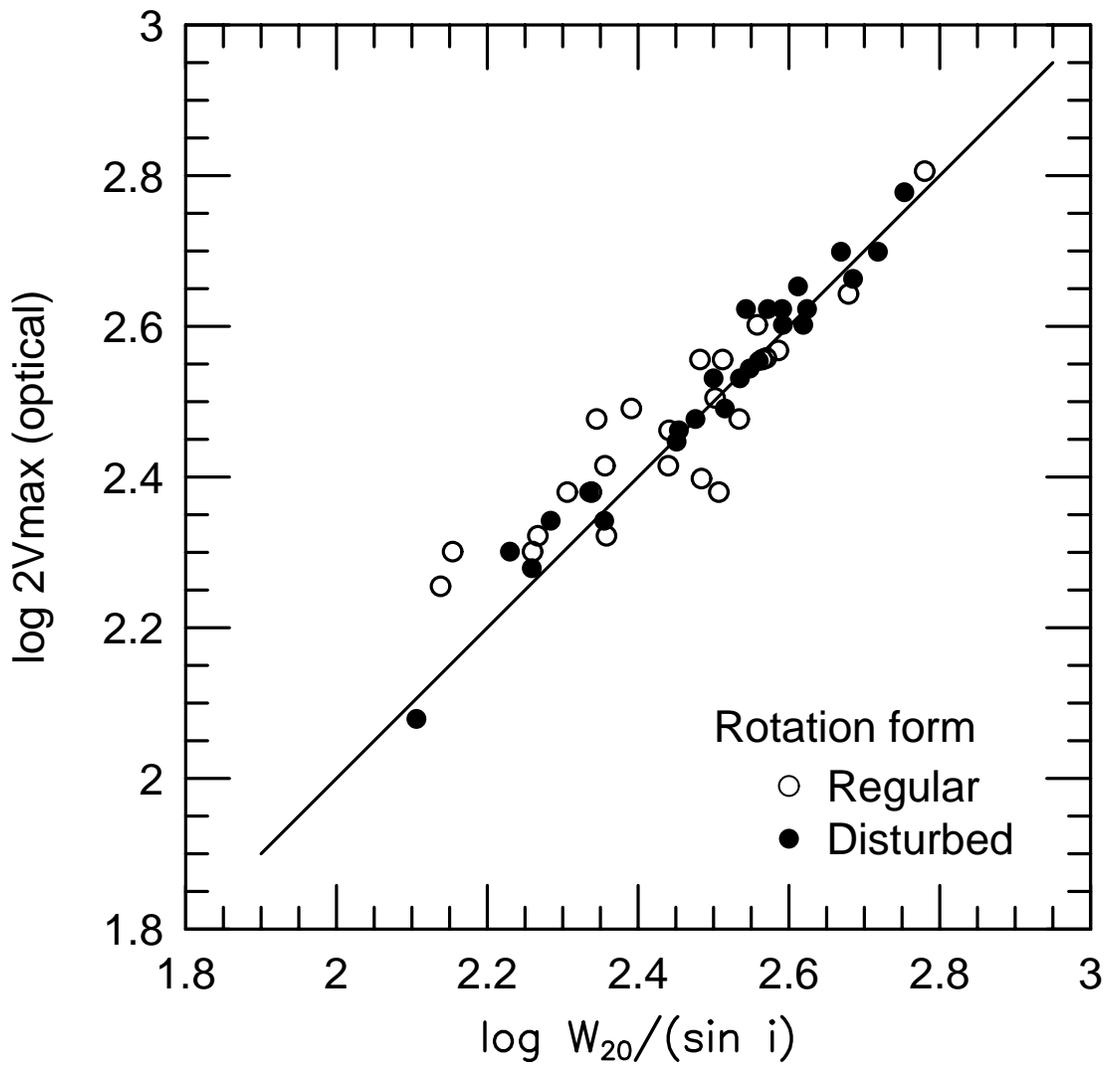
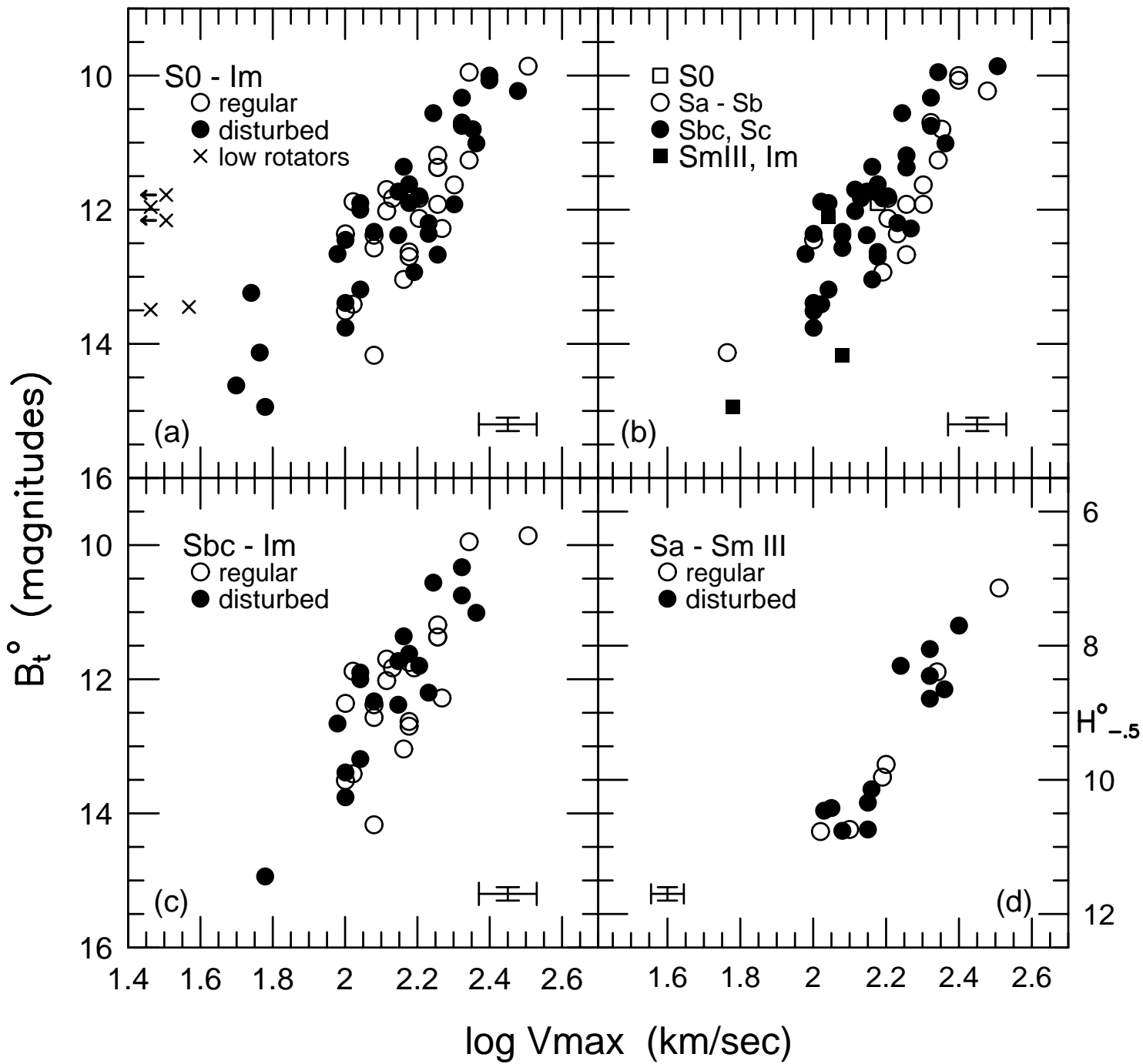
\end{document}